\documentstyle[12pt]{article}

\thicklines                         % thick straight lines for pictures (LaTeX)
\def\a{\alpha}

\def\d{\delta}
\def\f{\phi}                    %       \varphi
\def\g{\gamma}

\def\j{\psi}

\def\l{\lambda}
\def\m{\mu}

\def\p{\pi}                     % Also, \varpi
\def\th{\theta}                  %       \vartheta
\def\s{\sigma}                  %       \varsigma

\def\D{\Delta}

\def\G{\Gamma}
\def\J{\Psi}

\def\O{\Omega}

\def\S{\Sigma}

\def\cc{{\cal C}}

\def\ch{{\cal H}}   % overridden by cosh !!

\def\cl{{\cal L}}
\def\cm{{\cal M}}

\def\cbo{{\,\raise-.15ex\Sc [\,}}                       % curly "
\def\sl#1{\rlap{\hbox{$\mskip 1 mu /$}}#1}      % good slash for lower case
\def\Bar#1{\overline{#1}}                       % big bar
\def\ddt#1{{\buildrel {\hbox{\LARGE .\kern-2pt.}} \over {#1}}}% double dot-over
\def\beqn#1{ \renewcommand{\theequation}{#1}
             \begin{equation} }
\def\eeqn{\end{equation}}
\def\beqr#1{ \renewcommand{\theequation}{#1}
             \begin{eqnarray} }
\def\eeqr{\end{eqnarray}}
\def\NON{\nonumber\\}
\def\beqrabc#1{ \setcounter{equation}{0}
                \renewcommand{\theequation}{#1\alph{equation}}
                \begin{eqnarray} }
\def\beqrn#1#2{ \setcounter{equation}{#2}
                \renewcommand{\theequation}{#1.\arabic{equation}}
                \begin{eqnarray} }

\def\APH#1{Ann.~Phys.~{\bf #1}}
\def\CMP#1{Comm.~Math.~Phys.~{\bf #1}}
\def\IJMP#1{Int.~J.~Mod.~Phys.~{\bf #1}}

\def\NPB#1{Nucl.~Phys.~{\bf B#1}}
\def\NPBP#1{Nucl.~Phys.~(Proc.~Suppl.) {\bf B#1}}
\def\PLB#1{Phys.~Lett.~{\bf B#1}}
\def\PRD#1{Phys.~Rev.~{\bf D#1}}
\def\PR#1{Phys.~Rev.~{\bf #1}}
\def\PRL#1{Phys.~Rev.~Lett.~{\bf #1}}

\def\APP#1{Acta.~Phys.~Pol.~{\bf B#1}}

\def\sstyle{\scriptstyle}

\def\rhs{\mbox{r.h.s.} }

\def\ie{\mbox{i.e.} }
\def\eg{\mbox{e.g.} }

\def\frac#1#2{ {\sstyle {#1\over #2} } }
\def\det#1{{\rm det}\left(#1\right)}

\def\hc{{\rm h.c.\,}}

\def\half{{1\over 2}}

\def\ch{\,{\rm cosh}}
\def\sh{\,{\rm sinh}}

\def\det{{\rm det}}

\def\Dtp{\O_+}
\def\Dtm{\O_-}
\def\Dhp{\hat\O_+}
\def\Dhm{\hat\O_-}
\def\nbl#1{\nabla\!_#1 \,}

\begin{document}
\noindent  \hfill WIS--93/20/FEB--PH
\par
\begin{center}
\vspace{15mm}
{\large\bf Chiral Fermions from Lattice Boundaries}\\[5mm]
{\it by}\\[5mm]
Yigal Shamir\\
Department of Physics\\
Weizmann Institute of Science, Rehovot 76100, ISRAEL\\[15mm]
{ABSTRACT}\\[2mm]
  \end{center}
\begin{quotation}
  We construct a model in which four dimensional chiral fermions
arise on the boundaries of a five
dimensional lattice with free boundary conditions in the fifth direction.
The physical content is similar to Kaplan's model of domain wall fermions,
yet the present construction has several technical advantages.  We discuss
some aspects of perturbation theory,  as well as possible applications of
the model both for lattice QCD and for the on-going attempts to construct
a lattice chiral gauge theory.

\end{quotation}

\vspace{30mm}
\noindent email: ftshamir@weizmann.bitnet

\newpage

\noindent {\bf 1.~~ Introduction}
\vspace{3ex}

  The Electro-Weak sector of the Standard Model is a chiral gauge theory. This
fact motivated numerous attempts to construct a lattice model whose continuum
limit will be a chiral gauge theory~[1-5]. (See ref.~[6] for a review).

  The basic stumbling block which was recognized already in Wilson's
original work~[7] is the doubling problem. Following earlier work by
Karsten and Smit~[8], Nielsen and Ninomiya~[9] proved that species
doubling (and hence a vector-like spectrum) is an unavoidable consequence of
any free fermion lattice hamiltonian which satisfies some mild assumptions.
Thus, attempts of construct chiral lattice gauge theories fall into two
classes.
One approach is to obtains a chiral spectrum at tree level
by giving up one of the assumptions necessary of the validity of the no-go
theorem\footnote{
 Historically, the work of Drell, Weinstein and Yankielowicz~[1]
preceded and, to a large extent, motivated the work of refs.~[8]
and~[9].}~[1,2].
 Typically, the assumption which is violated is that the
dispersion relation has a continuous first derivative. In the second approach
the quadratic hamiltonian has a vector-like spectrum, and one tries to
eliminate the doublers dynamically by introducing some strong
interactions which are more effective for the doublers~[3-5].
However, until now, no lattice gauge model whose continuum limit is
consistent with the requirements of Lorentz invariance, unitarity etc.
has been shown to have a chiral spectrum~[10-12].

  Recently, Kaplan~[13] proposed that a four dimensional, chiral lattice gauge
theory may be constructed if one starts with a theory of
massive Dirac fermions in {\it five} dimensions,
provided the fermion mass has the shape of a domain wall.
Restricted to the fifth direction, the free fermionic hamiltonian
has a zero mode with definite chirality localized on the domain wall.
{}From the point of view of the four dimensional domain wall, this zero mode is
a chiral fermion.

  For a range of values of the Dirac mass $M$ and the Wilson parameter $r$,
the chiral fermion has no doublers if the fifth direction is infinite. But
when the fifth direction is taken to be finite a doubler appears~[13-15].
For example, if we choose periodic boundary conditions
an anti-domain wall  must appear somewhere, and the chiral
fermion on the anti-domain wall has the opposite chirality.

  As long as one considers the coupling of the fermions to an external gauge
field, this gauge field can be taken to be five dimensional. But at the
dynamical level, the introduction of a five dimensional gauge field requires
one
to deal with a host of highly non-trivial issues before the existence of a
non-trivial continuum limit of any sort can be
established. We will not elaborate on this possibility in this paper.

  The approach we adopt is to consider the fifth direction as a sophisticated
flavour space, and not as a full fledged additional space coordinate.
The interacting theory is defined by coupling the fermions to a {\it
four dimensional} gauge field~[16,17]. This  means that the link variables
satisfy $U_\m(x,s)=U_\m(x)$, $\m=1,\ldots,4,$ independently of $s$,
that $U_5(x,s)=1$
and that the gauge field action is any four dimensional lattice action which
reduces to the standard continuum action in the classical continuum limit.
Here $x_\m$  are the usual four dimensional coordinates and
$s$ denotes the fifth direction. In the case of a finite lattice we denote the
number of sites in the fifth direction by $N$.

  The first question is, are we dealing with a chiral or
a vector-like lattice theory at {\it tree level}.
The Nielsen-Ninomiya theorem asserts that,
assuming  (a) complex field formulation,
(b)  hermiticity of the hamiltonian and (c)  continuous first derivative
of the dispersion relation,   the massless spectrum of free fermions on
a regular lattice must consist of an equal
number of positive helicity (``right handed'') and negative helicity
(``left handed'')
Weyl fermion. (One can replace assumption (c) by the {\it sufficient}
conditions
that (c1) the number of  fermionic degrees of freedom per
(four dimensional) site is
finite and (c2) the hamiltonian has a short range. These conditions ensure that
the only possible singularities in the dispersion relation are level
crossings).
Moreover, (d) given any charge which is exactly conserved, locally defined and
has discrete eigenvalues, the equality of the number of left handed and
right handed fermions holds in every charged sector separately.

  At the level of a free fermion theory, the relevant charges in Kaplan's model
are the generators of the non-abelian symmetry that we intend to gauge.
If the fifth direction is infinite, the tree level spectrum
can be chiral because assumption (c1) is violated. Specifically, for a fixed
four momentum one has an effective one dimensional lattice problem, whose
spectrum consists of a continuous part as well as possible bound states.
When the energy of the bound state reaches the continuum threshold, the bound
state disappears from the spectrum, giving rise to a singularity in the
dispersion relation~[13-16].

  Narayanan and Neuberger~[16] considered the introduction of
gauge fields directly in the ``infinite fifth direction setting''.
To set up perturbation theory, they  define a {\it four dimensional} current
$J_\m(x)=\sum_{s=-\infty}^{\infty}j_\m(x,s)$.
Here $j_\m(x,s)$ stands for the first four components of the five dimensional
current. Apart from an unimportant substruction related to the presence of an
infinite number of heavy flavours, perturbation theory is defined in the usual
way in terms of correlators of any finite number of currents.

  As long as one is interested in perturbative results one can work
directly with the Feynmann rules of ref.~[16].
But, in order to properly define the model at the
non-perturbative level, one must consider a sequence of lattice
theories with a {\it finite} fifth direction such that, in the limit
$N\to\infty$, the Feynmann rules of ref.~[16] are reproduced
in weak coupling perturbation theory.
Since condition (c1) in now fulfilled, if the charged fermions spectrum is
to remain chiral in these finite lattice models, another assumption of
the Nielsen-Ninomiya theorem
must be violated. Assumptions (a) and (c2) are inherent properties of the
domain wall model. Consequently, either hermiticity of the hamiltonian
or tree level gauge invariance (or both) will have to be sacrificed.
Either way unitarity is violated, as pointed out already by the authors of
ref.~[16]. For this approach to succeed one must construct a specific
sequence of lattice models with the above properties, and show that
all violations of unitarity tend to zero in the limit (provided the spectrum
is anomaly free in the usual sense).

  If one insists on hermiticity and gauge invariance of the
finite lattice action, all
the conditions of the Nielsen-Ninomiya theorem are fulfilled and so the lattice
theory is vector-like at tree level~[13,14,17].
The question then is whether one can
decouple the doublers dynamically and achieve a chiral spectrum in the
continuum limit of the interacting theory. The hope is that the
specifics of this model will be sufficiently different from previous
models which have failed to
produce a chiral spectrum~[3-5]. Whether or not this is the case
is the  subject of on-going investigations.

   A characteristic property of all versions of the domain wall model is
that as the number of sites in the fifth direction grows,
all correlation functions tend to a limit like $e^{-N}$. The limiting
behaviour is therefore achieved for relatively small $N$. But it
remains to be seen whether a consistent chiral gauge theory can
be achieved in the continuum limit of some version of the model.

  In this paper we have little to say regarding the difficult question
of maintaining simultaneously a chiral spectrum and a consistent,
interacting gauge theory in the continuum limit. Our purpose is to
eliminate some unessential technical complications present in
Kaplan's original model, as well as to continue the
study of the model at the level of perturbation theory.

  In the first part of this paper we construct a five dimensional model
in which four dimensional
chiral fermions arise on the boundaries of a five dimensional slab with
free boundary conditions in the fifth direction.
As in the domain wall model, if the fifth direction is
(semi)-infinite there is a single chiral fermion on the four dimensional
boundary. But when the fifth direction is taken to be finite, a doubler
appears on the other boundary.

  It is well known that the boundary
of a sample can have its own dynamics, described by an effective
field theory in one less dimension.
In particular, one finds an intimate relation between Chern-Simons terms
in odd dimensions and anomalies in even dimensions~[18,19].
This relation plays a central role in Kaplan's model~[13,20]. These ideas  have
also found interesting applications in the context of the Quantum Hall
Effect~[21].

  The boundary fermions model is really a variant of Kaplan's model.
In fact, a careful examination of refs.~[13-15]  reveals that the zero mode's
spectrum is always determined by requiring normalizability of the wave
function on the side where $M$ and $r$ have the same sign. On the other
side of the domain wall this condition is always fulfilled.
If we ``discard'' that
side of the wall we arrive at the boundary fermion model.

  The present construction has  several technical advantages which merits
its discussion separately. Analytical expressions, in particular the
propagator, take a much simpler form in the boundary fermion scheme.
In the domain wall model, the propagator for an infinite fifth direction was
calculated in ref.~[16]. Here we give the propagator both for the semi-infinite
and the
finite lattice cases. Also, in numerical simulations one should obtain the same
accuracy in the boundary fermion model by taking half as many sites in
the fifth direction. This is simply because the fifth direction must extend
only on one side of the four dimensional boundary.

  In the second part of this paper we discuss the introduction of a
four dimensional gauge field. We first consider the use of the
model for lattice QCD.
When slightly modified, the model contains an additional parameter $m$
which in QCD plays the role of the current mass.
Unlike the case of Wilson fermions, we show that perturbative correction
to the quark mass are proportional to $m$. Consequently,
chiral perturbation theory is valid, and the
model can be used to study chiral symmetry breaking in lattice QCD.

  We next discuss  a ``mirror fermion'' model.
In order to construct the model we
introduce two five dimensional slabs, one for the charged fermions and
one for the neutral fermions. We also introduce a charged scalar (Higgs)
field. We show that in the broken symmetry phase, the model can naturally
lead to a large hierarchy between mirror fermions and ordinary fermions masses.
However, there are difficulties, and it is not clear whether the model can have
a continuum limit describing an {\it interacting} chiral theory.

  This paper is organized as follows. In sect.~2 we discuss the
boundary fermion model
on a semi-infinite lattice. In sect.~3 we discuss the model on a finite
lattice.
In sect.~4 we consider the introduction of gauge fields. Sect.~5 contains some
concluding remarks regarding the prospects of obtaining an interacting chiral
gauge theory in the continuum limit of domain wall or boundary fermions models.

\newpage
\noindent {\bf 2.~~ Boundary fermions on a semi-infinite lattice}
\vspace{3ex}

  We begin with the free field theory defined
on a semi-infinite five dimensional
lattice. For simplicity we work in five dimensions, but everything generalizes
to other odd dimensions as well. We mainly work with four
dimensional momentum eigenstate and disregard possible finite size effects
in the usual four dimensions. The properties of the boundary fermion model
presented in this section are similar to those of the domain wall
model~[13-16], and we rederive them
here for the convenience of the reader. Of course, technical detail such as
the explicit form of the bound states and the propagator are different (and
simpler) in the boundary fermion model.

  the action is given by $\cl=\cl_4 +\cl_5$ where $\cl_4$ is the usual
four dimensional Wilson action
summed over all $s$, and $\cl_5$ contains the couplings in the fifth direction.
The lattice spacing is taken to be $a=1$. Explicitly
$$ \cl_4 = \sum_{x,s,\m}\Bar\j(x,s)\g_\m\partial_\m\j(x,s) +
         M\sum_{x,s}\Bar\j(x,s)\j(x,s) + {r\over 2} \cl_W \,,\eqno(1)$$
$$ \cl_W = \sum_{x,s,\m}\Bar\j(x,s)\nbl\m\j(x,s)\,. \eqno(2)$$
$r$ is the Wilson parameter. We will mainly work with $r=1$ and only briefly
discuss the case $r\ne 1$. The lattice difference operators are defined by
$$ \partial_\m\j(x,s) = \half(\j(x+\hat\m,s) - \j(x-\hat\m,s)) \,, \eqno(3)$$
$$ \nbl\m\j(x,s) = \j(x+\hat\m,s)+\j(x-\hat\m,s)-2\j(x,s) \,. \eqno(4)$$
$\cl_5$ splits into a sum over all $s>0$ denoted $\cl'_5$ and a boundary term
$\cl^0_5$, where
$$ \cl'_5= \sum_{x,s>0}\Bar\j(x,s)\g_5\partial_5\j(x,s) +
         {r\over 2}\sum_{x,s>0}\Bar\j(x,s)\nbl{5}\j(x,s)  \,,\eqno(5)$$
$$ \cl^0_5 = \half\sum_{x}\Bar\j(x,0)\g_5\j(x,1) +
           {r\over 2}\sum_{x}\Bar\j(x,0)(\j(x,1)-2\j(x,0)) \,. \eqno(6)$$
Notice that $\cl^0_5$ is obtained from eq.~(5) by setting $s=0$
and dropping terms containing fields at the non-existing sites with $s=-1$.

  We now go to momentum eigenstates and set $r=1$. The action becomes
$$ \cl = \int_p \sum_{s,s'} \bar\j(-p,s) D(s,s';p) \j(p,s') \,, \eqno(7)$$
where the integral is over the Brillouin zone, and
$$  D(s,s';p) = \th(s)\th(s') D_0(s,s';p) \,.\eqno(8)$$
Here $\th(s)=1$ for $s\ge 0$ and $\th(s)=0$ for $s<0$. $D_0(s,s';p)$ is the
Dirac operator on an infinite fifth direction
$$ D_0(s,s';p) = \half(1+\g_5)\d_{s+1,s'} +\half(1-\g_5)\d_{s-1,s'} -
           (b(p) + i \sl\bar{p})\d_{s,s'} \,,
\eqno(9)$$
$$ \bar{p}_\m=\sin p_\m \,,$$
$$ b(p)= 1-M+\sum_{\m} (1-\cos(p_\m)) \,.\eqno(10)$$

  We now take $M$ to lie in the interval $0<M<2$. (Later we will see that one
must further restrict $M$ to lie in the interval $0<M<1$).
The physical content of the model is more transparent if we consider the
lattice
{\it hamiltonian}
$$ H(s,s';p_k) =  \g_4 D(s,s';p_k,p_4=0)\,. \eqno(11)$$
Notice that in the hamiltonian framework, the fifth direction is really
a forth space coordinate with a semi-infinite range.
We will nevertheless keep calling it the ``fifth direction''.
We will show that the spectrum of $H(s,s';p_k)$ contains a right handed Weyl
fermion that lives on the space boundary. The Weyl fermion is described by
a bound state with energy $E^2=\bar{p}^2_k$, which $H(s,s';p_k)$ admits
for every $p_k$ in the
domain $|b(p_k)|<1$. Notice that, for our choice of $M$, the domain in which
chiral fermions exist contains the origin of the Brillouin zone but no points
in which some of the momentum components are equal to $\p$.
 As a result, the chiral fermion has no doublers~[13].
(This situation will change when we make the fifth direction finite).

  The bound state wave function has the form
$$ \J_R(s,p_k)=U(s)(1+\g_5)\j(p_k) \,, \eqno(12)$$
where the right handed spinor $\j(p_k)$ is a helicity eigenstate
$$ \sum_{j=1}^3 \s_j \bar{p}_j \, \j(p_k)=E\j(p_k) \,.\eqno(13)$$
Substituting this into the eigenvalue equation $H \J_R = E \J_R$ we find that
$U(s)$ must satisfy $ U(s+1) = b(p_k) U(s)$. The normalized solution is
$$ U(s) = (1-b^2(p_k))^\half \, b^s(p_k) \,.\eqno(14)$$
As promised, this solution is normalizable provided $|b(p_k)|<1$. As $|b(p_k)|$
approaches one, the solution decreases slower and slower, until at $b(p_k)=1$
it becomes a continuum eigenstate with vanishing fifth component of momentum.

  Before we turn to the propagator let us briefly discuss the
continuous spectrum.
Because of complete reflection at the boundary, the
continuum eigenstates are standing waves in the fifth direction.
Denoting the fifth component of the momentum by $p_5$,
the energy of a continuum eigenstate is
$E^2 =  (b(p_k)-\cos p_5)^2 + \bar{p}_k^2+\bar{p}_5^2$.
If the three-momentum $p_k$ is fixed, the minimal energy
is obtained for $p_5=0$ and is given by
$\min |E|=((b(p_k)-1)^2+\bar{p}^2_k)^{1/2}$. We see that $\min |E|$ coincides
with the energy of the bound state on the surface define by $b(p_k)=1$, which
is
recognized as the boundary of the domain in which the chiral fermion exists.
The bound state disappears when its energy reaches the continuum threshold.
This completes our discussion of the spectrum of the lattice hamiltonian.

  We now return to the Dirac operator~(8) of the eculidean formulation.
Unlike the Hamiltonian, the eigenstates of the euclidean Dirac operator do
not have simple physical interpretation. The fact that the Dirac operator is
complex allows funny things to happen. In particular, the euclidean
Dirac operator has a bound state only for $p_\m=0$~[22]. On the other hand,
the second order operators $D D^\dagger$ and $D^\dagger D$ are hermitian  and
non-negative, and so their spectrum is perfectly well
behaved. When we speak about a bound state spectrum in euclidean space we will
always refer to one of the second order operators.

  Inspite of the peculiar spectrum of $D(s,s';p)$, the fermionic
{\it propagator}  $G_F$ does not show any unexpected behaviour, because
$G_F=D^\dagger G$ where $G$ is the propagator of the second order
operator $D D^\dagger$. As a first stage towards the construction of $G_F$,
we consider the Dirac operator $D_0$
defined on an infinite $s$ direction (eq.~(9)). Going to the second order
operator $D_0 D_0^\dagger$, it is easy to check that
the two homogeneous solutions are given by $\exp(\pm\a s)$, where $\a$ is
defined by the positive solution of the equation~[16]
$$ 2\ch\a(p)={1+b(p)^2+\bar{p}^2\over b(p)} \,. \eqno(15)$$
Notice that if $b(p)$ has a zero then $\a(p)$ has a logarithmic singularity.
To avoid such singularities we must take $0<M<1$.
(There is in fact a second allowed range, given in $2n+1$ dimensions by
$4n+1<M<4n+2$. In this case the chiral fermion  occurs near the corner (rather
than near the origin) of the Brillouin zone~[15]. The physics
in both cases is the same, and so we always assume $0<M<1$).
The \rhs of eq.~(15) is always greater that two. This ensures that
$\a(p)$ is an analytic function in the entire Brillouin zone. Notice that
both $b(p)$ and $e^{-\a(p)}$ tend to $1-M$ for $p_\m\to 0$.

The inverse $G_0$ of the second order operator $D_0 D_0^\dagger$ is given by
$$ G_0(s,s')=Be^{-\a|s-s'|} \,,\eqno(16)$$
$$ B^{-1}=2b\sh\a \,.\eqno(17)$$
The two chiralities are decoupled in the second order operators.
In particular
$$ D D^\dagger = \half(1+\g_5)\Dtp + \half(1-\g_5)\Dtm
\,,\eqno(18)$$
where $\Dtp$ and $\Dtm$ carry no Dirac indices.
This implies a similar decomposition for the propagator
$$G = \half(1+\g_5)G_+ + \half(1-\g_5)G_- \,.\eqno(19)$$

  We begin with the construction of $G_-(s,s')$. To this end,
we apply $\Dtm(s'',s)$ to $G_0(s,s')$
and check what is the deviation of the result from $\d_{s'',s'}$. Since
$D_0 D_0^\dagger$ and $\Dtm$ contain only nearest neighbour
coupling, the deviation vanishes unless $s''=0$. For $s''=0$ one has
$$ \sum_{s\ge 0} \Dtm(0,s)\, G_0(s,s')-\d_{0,s'} =
   Be^{-\a s'}(be^{-\a}-1) \,.
\eqno(20)$$
Similarly, for the homogeneous solutions we find
$$\sum_{s\ge 0} \Dtm(s'',s)\, e^{\pm\a s} = \left\{ \begin{array}{ll}
     0\,, & s''>0\,, \\
     b\, e^{\mp\a} -1 \,, & s''=0 \,.
\end{array}\right. \eqno(21)$$
Eqs.~(20) and (21) suggest that $G_-(s,s')$ has the form
$$ G_-(s,s') = G_0(s,s') + K(s')\, e^{-\a s} \,.\eqno(22)$$
That is to say, each column of $G_-$ can be constructed from the corresponding
column of $G_0$, plus a column vector proportional to a homogeneous solution.
In order that $G_-(s,s')$ satisfy physical boundary conditions we allow only
the exponentially decreasing solution on the \rhs of eq.~(22). Since
the second order operators (and hence their propagators) are symmetric,
we must have
$K(s')=A_- \exp(-\a s')$ where the amplitude $A_-$ depends on $p_\m$.
The amplitude $A_-$ is easily found using
eqs.~(20) and~(21). Following the same steps we construct also $G_+$.
The final result is
$$ G_\pm(s,s') = G_0(s,s') + A_\pm \, e^{-\a(s+s')}\,,
\quad\quad s,s'\ge 0 \,, \eqno(23)$$
$$ A_- = B\,e^{-2\a}\,{e^{\a}-b \over b-e^{-\a}} \,,\eqno(24)$$
$$ A_+ = -B\,e^{-2\a} \,.\eqno(25)$$

  Let us now consider the physical content of these equations. While
the amplitudes $B$ and $A_+$ are regular for all values of $p_\m$, the
amplitude $A_-$ is singular for small $p_\m$,
$$ A_-(p) = {M(2-M)\over p^2}+ \mbox{regular terms}\,, \quad\quad p_\m \to 0
\,.\eqno(26)$$
This singularity reflects the existence of a bound state whose eigenvalue
tends to zero for $p_\m \to 0$. The eigenvalue $\l_0^2$ of this bound state
can be read off the propagator as follows. Using eq.~(14) for small $p_\m$,
the contribution of the bound state to the propagator is approximated by
$\l_0^{-2} M(2-M) (1-M)^{-s-s'}$. Equating this to the second term on the \rhs
of eq.~(23) and using eq.~(26) we find $\l_0^2 = p^2$ as expected.
Although this time the answer was known beforehand, we have explained this
trick for extracting a vanishingly small eigenvalue from the propagator
because it will be useful to us again later.

  We finally notice that the $1/p^2$ singularity in $G_-$ gives rise to a
chiral pole in the fermionic propagator
\beqr{27}
  G_F & = & D^\dagger G \NON
      & = & {i\over 2}(1+\g_5) {M(2-M)\over \sl{p} }(1-M)^{-s-s'}
             + \mbox{regular terms}\,.
\eeqr

  As already mentioned in the introduction, the dispersion relation is singular
on the surface $b(p)=1$ where the bound state disappears from the spectrum.
Interestingly, the propagator does not show any singularity at $b(p)=1$.
This is because the normalization factor of the bound state tends to zero as
$b(p)$ approaches one, and so the contribution of the bound state to the
propagator vanishes for fixed $s$ and $s'$. One must check, however,
whether or not some sort of singularity reappears in perturbation theory when
summations over the fifth direction are carried out.

\vspace{5ex}
\noindent {\bf 3.~~ Boundary fermions on a finite lattice}
\vspace{3ex}

  We now proceed to discuss the model with a finite fifth direction
$0 \le s\le N$. The crucial difference is that now
a second chiral fermion appears on the new boundary at $s=N$.
Not surprisingly, this fermion has the opposite chirality from
the one at $s=0$.  Strictly speaking, there is a tiny mixing between the
two chiral modes, which vanishes like $(1-M)^N$. Such exponentially small
modifications will cause us no concern and henceforth we neglect them.

  The propagator can be constructed using the same method as before.
This is convenient because the information about the low energy
excitations is encoded in the singularities of the propagator, and, in any
event, what is needed to develop perturbation theory is the propagator.

  Denoting quantities that belong to the finite lattice model by a hat,
the Dirac operator is now given by
$$ \hat{D}(s,s')= \th(N-s)\th(N-s')D(s,s') \,.\eqno(28)$$
Before we actually construct the propagator, it is useful to consider a
certain generalization of the model. We notice that if we start from
the Dirac operator defined on a finite fifth direction which has the topology
of a {\it circle}, then $\hat{D}$ can be obtained by cutting the link
connecting the sites $s=0$ and $s=N$. Now, when $M$ is constant,
using a fifth direction with the
topology of a circle (or, equivalently, imposing periodic boundary conditions)
gives rise to no low energy excitations. Therefore, if starting from the Dirac
operator of eq.~(28) we gradually turn on the link connecting the sites $s=0$
and $s=N$, we expect that the two Weyl fermions will form a Dirac fermion
whose mass is proportional to the strength of this link.

  We are therefore lead to consider the Dirac operator
$$ \hat{D}(s,s';m) = \hat{D}(s,s')
                     + {m\over 2}(1-\g_5)\d_{s,1}\d_{s',N}
                     + {m\over 2}(1+\g_5)\d_{s,N}\d_{s',1}\,.
\eqno(29)$$
As we will see, up to a constant, $m$ indeed plays the role of a Dirac mass for
the light fermions. Moreover, unlike the case of Wislon fermions,
perturbative correction will always be proportional to $m$, \ie chiral
perturbation theory is valid. Thus, the model can be used to
study chiral symmetry breaking in lattice QCD. (Whether this advantage merits
the extra trouble involved in going to a five dimensional setting is a
practical question that we will not address here).

  We now proceed to construct the propagator $\hat{G}_F$ of the Dirac
operator $\hat{D}$ for a general value of $m$. The propagator for the massless
case will be obtained by simply setting $m=0$. As before,
$\hat{G}_F = \hat{D}^\dagger \hat{G}$ where $\hat{G}$ is the propagator of the
second order operator  $\hat{D}\hat{D}^\dagger$.
The two chiralities again decouple in the second order operators.
Furthermore, we  now have
$$ \Dhp(s,s')=\Dhm(N-s,N-s') \,. \eqno(30)$$
Thus, $\hat{G}_+(s,s')$ is obtained from $\hat{G}_-(s,s')$ by the
replacement $s,s'\to N-s,N-s'$. The notation is the same as in
eqs.~(18) and~(19).

  The propagator must have the following form
\beqr{31}
   \hat{G}_-(s,s') & = & G_0(s,s') + \hat{A}_- \, e^{-\a(s+s')} +
       \hat{A}_+ \, e^{-\a(2N-s-s')} \NON
       & & +  \hat{A}_m(e^{-\a(N+s-s')}+e^{-\a(N+s'-s)})\,.
\eeqr
This time, the deviation of $\sum_s\hat{D}\hat{D}^\dagger(s'',s) G_0(s,s')$
from $\d_{s'',s'}$ vanishes except for $s''=0$ and $s''=N$. The same is
true for $\sum_s \hat{D}\hat{D}^\dagger(s'',s) \exp{(\pm\a s)}$.
Thus, in order to construct the $s'$-th column of the
propagator we need a linear combination of $G_0(s,s')$ and both of the
homogeneous solutions. Taking into account the symmetry of the propagator we
arrive at eq.~(31).

  In solving for the $s'$-th column of the propagator, the ($s'$-dependent)
coefficients of $\exp{(\pm\a s)}$ are two unknowns which are determined by
solving a two by two matrix equation. There are actually two such equations
in which $\hat{A}_m$ appears twice,
because the $s'$-dependence of the coefficients can be either $\exp{(\a s')}$
or $\exp{(-\a s')}$. Explicitly
$$ \cc \left( \begin{array}{c} \hat{A}_- \\ \hat{A}_m \end{array} \right) =
      B  \left( \begin{array}{c} 1-b\,e^{-\a}-m^2 \\ mb \end{array} \right)\,,
\eqno(32)$$
$$ \cc \left( \begin{array}{c} \hat{A}_m \\ \hat{A}_+ \end{array} \right) =
      B  \left( \begin{array}{c} mb \\ -b\,e^{-\a} \end{array} \right)\,,
\eqno(33)$$
where
$$ \cc = \left( \begin{array}{ccc}
           b\,e^\a+m^2-1 & & -mb \\  -mb & & b\,e^\a
         \end{array} \right)\,.
\eqno(34)$$
the solutions are
$$ \hat{A}_- = \D^{-1}B(1-m^2)(e^\a-b)\,, \eqno(35)$$
$$ \hat{A}_+ = \D^{-1}B(1-m^2)(e^{-\a}-b)\,, \eqno(36)$$
$$ \hat{A}_m = 2\D^{-1} B b m \ch\a \,,\eqno(37)$$
where
$$ \D=b^{-1}\det\,\cc= e^\a (b\,e^\a-1)+m^2(e^\a-b) \,.\eqno(38)$$

  In the limit $m=0$ we find $\hat{A}_m=0$, $\hat{A}_-=A_-$ and
$\hat{A}_+=A_+$.
Notice that $\cc$ is diagonal for $m=0$, which means that the contributions
to the propagator from the two boundaries are decoupled. (Strictly speaking,
at $m=0$ one is left with exponentially small off-diagonal terms in $\cc$,
which can be ignored as long as $p_\m$ itself is not exponentially small.
The infinite (four dimensional) volume limit can always be taken in such a way
that exponentially small four momenta never occur. All that is needed is to
take $N$ to be slightly bigger that the logarithm of the number of sites in
the ordinary directions).

  The interpretation of $m$ as current mass suggests that we should consider
the limit where both $m^2$ and $p^2$ are small in lattice units.
Again, the only
amplitude which is singular in this limit is $\hat{A}_-$. We can use the method
described earlier to extract the smallest eigenvalue from the singular part
of the propagator. The result is
$$ |\l_0|^2 =p^2+m^2M^2(2-M)^2 \,.\eqno(39)$$
The current mass of the light Dirac fermion is therefore $mM(2-M)$.
As promised, it is proportional to $m$.

\newpage
\noindent {\bf 4.~~ The interacting theory}
\vspace{3ex}

  We now proceed to discuss the interacting theory. We first consider the
vector-like model, which consists of a single five dimensional slab of charged
fermions, coupled to a four dimensional
gauge field as described in the introduction.
This model can be used to describe lattice QCD, and it
poses no conceptual difficulties. Its interesting
feature is that the current mass $m$ gets only multiplicative renormalization.
The chiral limit is therefore achieved by letting $m$ tend to zero.
In particular, we expect that the pion will be massless in this limit.

In order to verify this picture we have to show that in the
limit $m\to 0$ the Dirac fermion remains massless to all orders in
weak coupling perturbation theory. This, in turn, is true provided the
inverse fermion propagator calculated up to $n$-th order
$$\G^{(n)}(s,s';p)=\hat{D}(s,s';p)+g^2 \S^{(1)}(s,s';p)+\ldots
    +g^{2n}\S^{(n)}(s,s';p)\,,\eqno(40)$$
has one zero mode on each boundary for $p_\m=0$. Here $\S^{(k)}(s,s';p)$
is the $k$-th order self energy.

  The physical reason for the stability of the zero mode is the following.
In weak coupling perturbation theory, corrections to the tree level inverse
fermion propagator are small. Moreover, since the five dimensional fermion
is massive, these correction are exponentially suppressed as $|s-s'|$ grows.
Thus, except for an exponentially small effect the zero modes
living on the  two boundaries
cannot mix with each other, and so a mass term cannot develop.

  A discussion of the perturbative stability of the (single) zero mode has been
given for the infinite lattice case in ref.~[16]. There, the stability was an
immediate
consequence of the absence of additional low energy states that could mix
with the  zero mode. Here we extend the analysis to the finite lattice case and
show that, thanks to exponential suppression of all correlations in the fifth
direction, the masslessness of the light fermions is maintained in spite of the
fact that the light spectrum is vector-like.

  Let us now examine this issue in some detail. The gauge boson
propagator is proportional to $\d_{s,s'}$, and so the one loop self energy is
diagonal in $s$-space $\S^{(1)}(s,s';p)=\d_{s,s'} \S^{(1)}(p)$.
The Dirac operator preserves its tree level structure and, for small $p_\m$,
the only change is in the five dimensional mass
$M\to M^{(1)}=M+g^2 \S^{(1)}(p=0)$. As long as $0< M^{(1)} < 1$ all
the qualitative statements regarding the tree level spectrum and tree level
propagator apply. In particular, one chiral fermion exists on each boundary.

  At two or higher loop level there appear non-diagonal contributions
coming from intermediate states of three or more massive fermions.
At the $n$-th order these contributions decay at least as fast as
$(1-M^{(n-1)})^{3|s-s'|}$. We have to show that $\G^{(n)}(s,s';0)$ has a
homogeneous solution which decreases exponentially for $s\ll N$.
This is a sufficient condition for the existence of an (approximate)
bound state with
exponentially small energy near the boundary $s=0$. (A similar
statement applies for the other boundary).

  In order to see that the existence of such a homogeneous solution is stable
against any small modification of the tree level Dirac operator, it is
instructive to first check what happens if we take the Wilson parameter $r$
to be close to, yet different from $1$.
Consider the Dirac operator defined on an
infinite $s$-direction for $r\sim 1$. One can easily check that
for $p_\m=0$ this Dirac operator has {\it two} homogeneous solution for
each chirality. In the positive chirality sector,
one solution behaves approximately like $(1-M)^s$.
This is the only solution which is present for $r=1$. But for $r\ne 1$ there is
a second solution which behaves like $\left({1-r\over 1-M}\right)^s$.
When we restrict our system to a semi-infinite $s$-direction, the boundary
term in the Dirac operator picks a linear combination of the two solutions.
Since both solutions are normalizable for $s\ge 0$, the result is a
normalizable
solution too. As $r\to 1$ the second solution tends to
zero, until at $r=1$ we are left only with the solution $(1-M)^s$.

  A similar mechanism works for $\G^{(n)}(s,s';0)$.
Imagine that we turn on the elements of $\G^{(n)}(s,s';0)$ one diagonal at a
time.
If we consider only $s$-values which are far from both boundaries, the
introduction of every new diagonal gives rise to a new homogeneous solution.
If the new diagonal is below the main diagonal, the new homogeneous solution
decreases for $s\ge 0$ and so it causes no problem. If the new diagonal
is above the main diagonal, the new homogeneous solution increases for
$s\ge 0$. But when we look for a linear combination of homogeneous solutions
which satisfies the equation $\sum_{s'}\G^{(n)}(s,s';0)\J(s')=0$ near $s=0$,
we find that the boundary terms in $\G^{(n)}(s,s';0)$ allow us
an additional free parameter.
With the extra free parameter we can find a linear combination which does
not include the unwanted new solution.

  This completes our discussion of the perturbative stability.
Of course, at the non-perturbative level we expect
that chiral symmetry breaking will take place, giving rise to the usual
spectrum of confining theories including in particular a massless pion in the
limit $m\to 0$.

\vspace{3ex}

  We now proceed to discuss the mirror fermion model. This model in obtained
by taking two five dimensional slabs, each described at tree level by the
Dirac operator of eq.~(29). We next introduce gauge fields only on one slab.
This slab will describe the charged fermions. The other slab describes neutral
fermions. The final step involves the introduction of a charged scalar field
and the addition to the action of a Yukawa term which couples the boundary
layer $s=N$ of the charged fermions to the boundary layer $s=0$ of the
neutral fermions
$$ \cl\,_{Yukawa}=y\sum_x \left( \f(x)\bar\j(x,N) \j^0(x,0) + \hc \right)\,.
\eqno(41)$$
Here $\j(x,s)$ denotes the charged fermions and $\j^0(x,s)$ denotes the
neutral fermions. The light spectrum consists
of one left handed and one right handed charged fermions which we denote
$\j_R(x)$ and $\j_L(x)$, as well as one left handed and one right handed
neutral fermions denoted $\j^0_R(x)$ and $\j^0_L(x)$. The right handed fermions
arise from the two $s=0$ boundaries and the left handed fermions arise from the
$s=N$ boundaries. We repeat this construction for every irreducible
representation of the gauge group.

  In the broken symmetry phase we consider the Dirac fermions
$$ \j_1=\left( \begin{array}{c} \j^0_R \\ \j_L \end{array} \right)\,,
\quad\quad
    \j_2=\left( \begin{array}{c} \j_R \\ \j^0_L \end{array} \right)\,.
\eqno(42)$$
The mass matrix is
$$ \cm \left( \begin{array}{c} \j_1 \\ \j_2 \end{array} \right) =
   \left( \begin{array}{ccc}
           yv & & m \\  m & & 0
         \end{array} \right)
   \left( \begin{array}{c} \j_1 \\ \j_2 \end{array} \right)\,.
\eqno(43)$$
Here $v$ is the Higgs VEV. The mass matrix $\cm$ is evidently of the seesaw
type. Furthermore, perturbative stability implies that the off-diagonal
terms proportional to $m$ are only multiplicatively renormalized as in the
QCD case. (The effect of quantum correction on the Higgs VEV is
more subtle. See \eg ref.~[23]).
In the limit $m\ll v$, the mass matrix $\cm$ describes a heavy
fermion of mass $yv$ and a light fermion of mass $m^2/yv$. In particular,
taking the limit $m\to 0$ we find that the massless spectrum contains a
charged right handed fermion and a neutral left handed fermion.

\vspace{5ex}
\noindent {\bf 5.~~ Prospects}
\vspace{3ex}

  The massless spectrum of the above mirror fermion model is therefore
chiral, but the trouble is that in order
to achieve this we had to break the gauge symmetry spontaneously. If no
special fine tuning is made, the gauge bosons mass and the Higgs mass will be
of the same order of magnitude as the mirror fermions mass. Thus, in the low
energy limit we obtain a theory of {\it non-interacting} chiral fermions.

   The main difference between the model described above and mirror fermion
models~[4] based on Wilson fermions is that no fine tuning is needed in order
to obtain an undoubled spectrum of light fermions.
But the real question is whether one can find a
version of this model such that at some point in the phase diagram one
has simultaneously light gauge bosons and light chiral fermions. Whether
models based on domain wall or boundary fermions can do better in this
regard than models based on Wilson fermions is not clear.

  Certain difficulties pertaining to lattice models containing Higgs
fields were pointed out by Banks and Dabholkar~[23].
Some of them can be dealt with by going to a model of
the Eichten-Preskill type~[5] or to a mixed model. Other difficulties
are relevant even for composite Higgs and are therefore generic.

  Most disturbing is the fact that in all the models discussed in the
literature~[3-5] there does not seem
to be a clear mechanism that will distinguish between anomalous and
non-anomalous theories. This is true even for the Eichten-Preskill model~[5].
In this model one makes sure that every {\it global symmetry} which is
broken by instanton effects in the continuum model will already be
broken explicitly by the lattice action.
But it is the dynamics which has to determine in what
phase the gauge symmetry will be realized.

  In the event that the undoubled spectrum is anomalous,
the 'tHooft consistency condition~[24] implies that the theory must be in the
Higgs phase. This is because in the ungauged model, the non-zero contribution
of the light fermions to the anomaly can only be cancelled by a Goldstone
boson. (The complete spectrum cannot give rise to an anomaly because in the
underlying lattice theory the charge is exactly conserved).
But the mechanism which induces the spontaneous breaking of global chiral
symmetries~[25] does not distinguish between would-be anomalous and
non-anomalous theories. Thus, it is not
impossible that spontaneous symmetry breaking is a price that must
{\it always} be paid in order to obtain an undoubled massless spectrum,
including in theories whose undoubled spectrum is non-anomalous. This concern
is particularly relevant since in all models one attempts to obtain a chiral
massless spectrum {\it before} the gauge interactions are turned on.
Indeed, recent results~[12] provide strong evidence that the
Eichten-Preskill model undergoes spontaneous symmetry breaking, and that the
spectrum in vector-like throughout the entire phase diagram.

  The crucial issue is whether in our candidate lattice model there is
a mechanism which forbids the existence of an interacting chiral continuum
limit and which is operative only in the absence of anomaly cancellation.
In Kaplan's original paper it was suggested that such a mechanism does exist
in the domain wall model. Namely, in the absence of anomaly cancellation
there exists a Goldstone-Wilczek current~[19] away from the
domain wall that should prevent the decoupling of the heavy fermionic
degrees of freedom from the light ones.

  The work of Narayanan and Neuberger~[16] represents an attempt to
exploit this mechanism at the level of perturbation theory.
But in order that this mechanism be operative non-perturbatively
in a well defined lattice model, we have to sacrifice either
hermiticity of the hamiltonian or tree level gauge invariance (or both).
Typically, the resulting unitarity violating effects will arise in a
region which is deep indise the five dimensional space, far from the
region which supports the chiral fermions. Consequently, most of the
unitarity violating effects will be
formally suppressed by positive powers of the lattice spacing. (This statement
should be true for the effective action of the gauge fields obtained by
integrating out the fermions). The only
unsuppressed effect which cannot be cancelled by counter-terms should be
the usual triangle anomaly. However, even if we arrange for cancellation
of the usual triangle anomaly, there is a serious danger that
additional, {\it finite} unitarity violating effects will survive
and destroy the consistency of the model, because positive powers of the
lattice spacing can be compensated by divergent loop integrals when we
integrate over the gauge fields.

  On the other hand, if one insists on hermiticity and gauge invariance
of the lattice action we do not see how the mechanism described above
can be operative. Thus, in our view, the prospects of obtaining an interacting
chiral gauge theory in the continuum
limit of models of the kind described above do not seem very promising.
But more work has to be done before a definite conclusion can be reached.
At the very least, since these models do not require any fine tuning to
maintain
the masslessness of the light fermions, they can help us focus on the real
issue. Namely, the feasibility of
maintaining simultaneously light gauge bosons and a light chiral spectrum.

%\vspace{5ex}
%\centerline{\rule{5cm}{.3mm}}

\vspace{5ex}
\centerline{\bf Acknowledgements}
\vspace{3ex}

  I thank A.~Casher and S.~Yankielowicz for discussions. This research was
supported in part by the Basic Research Foundation administered by the Israel
Academy of Sciences and Humanities, and by a grant from the United States --
Israel Binational Science Foundation.

\vspace{5ex}
\centerline{\bf References}
\vspace{3ex}
\newcounter{00001}
\begin{list}
{[~\arabic{00001}~]}{\usecounter{00001}
\labelwidth=1cm}

\item S.D.~Drell, M.~Weinstein and S.~Yankielowicz, \PRD{14} (1976) 487, 1627.

\item C.~Rebbi, \PLB{186} (1987) 200.

\item J.~Smit, \APP{17} (1986) 531. P.~Swift, \PLB{145} (1984) 256.
G.T.~Bodwin and E.V.~Kovacs, \NPBP{20} (1991) 546.

\item I.~Montvay, \PLB{199} (1987) 89; \NPBP{4}  (1988) 443.

\item E.~Eichten and J.~Preskill, \NPB{268} (1986) 179.

\item I.~Montvay, in {\it Lattice 91}, \NPBP{26} (1992) 57.
J.~Smit, \NPBP{17} (1990) 3.

\item K.G.~Wilson, in {\it New Phenomena in Sub-Nuclear Physics} (Erice, 1975),
ed. A.~Zichichi (Plenum, New York, 1977).

\item L.H.~Karsten and J.~Smit, \NPB{183} (1981) 103.

\item H.B.~Nielsen and M.~Ninomiya, \NPB{185} (1981) 20,
{\it Errata} \NPB{195} (1982) 541; \NPB{193} (1981) 173.

\item L.H.~Karsten and J.~Smit, \NPB{144} (1978) 536; \PLB{85} (1979) 100.
J.M.~Rabin, \PRD{24} (1981) 3218. M.~Campostrini, G.~Curci and A.~Pelissetto,
\PLB{193} (1987) 279.

\item M.F.L.~Golterman, D.N.~Petcher and  J.~Smit, \NPB{370} (1992) 51.
M.F.L.~Golterman and D.N.~Petcher, \NPBP{26} (1992) 483.
W.~Bock, A.K.~De, C.~Frick, K.~Jansen and T.~Trappenberg, \NPB{371} (1992) 683.
See also ref.~6.

\item M.F.L.~Golterman and D.N.~Petcher \NPBP{26} (1992) 486.
M.F.L.~Golterman, D.N.~Petcher and E.~Rivas, Wash. U. preprint HEP/92-80,
\NPB{ } to appear.

\item D.B.~Kaplan, \PLB{288} (1992) 342.

\item K.~Jansen, \PLB{288} (1992) 348.

\item K.~Jansen and M.~Schmaltz, {\it Critical Momenta of Lattice Chiral
Fermions}, UCSD/PTH 92-29.

\item R.~Narayanan and H.~Neuberger, {\it Infinitely Many Regulator Fields for
Chiral Fermions}, RU-92-58.

\item D.B.~Kaplan, talk given at the Lattice 92 conference.

\item B.~Zumino, W.~Yong-Shi and A.~Zee,  \NPB{239} (1984) 477.
E.~Witten, \CMP{121} (1989) 351. S.~Elitzur, J.~Moore, N.~Seiberg and
A.~Schwimmer, \NPB{326} (1989) 108. G.~Dunne and C.~Trugernberger, \APH{204}
(1990) 281.

\item J.~Goldstone and F.~Wilczek, \PRL{47} (1981) 986.
C.G.~Callan and J.A.~Harvey, \NPB{250} (1985) 427.

\item M.F.L.~Golterman, K.~Jansen and D.B.~Kaplan, {\it Chern-Simons
Currents and Chiral Fermions on the Lattice}, submitted to \PRL.

\item X-G.~Wen,, \IJMP{B6} (1992) 1711. M.~Stone, \APH{207} (1991) 38.
A.~Cappelli, G.~Dunne, C.~Trugenberger and G.~Zemba, CERN preprints
TH-6702/92, TH-6784/93, \NPB{ } to appear.

\item Y.~Shamir, {\it The Euclidean Spectrum of Kaplan's Lattice Chiral
Fermions}, Weizmann preprint WIS-92/97/12-PH.

\item T.~Banks and A.~Dabholkar, \PRD{46} (1992) 4016.

\item G.~'tHooft, in {\it Recent Developments in Gauge Theories}, eds.
G.~'tHooft et al. (Plenum, New York, 1980), p.~135.

\item Y.~Nambu and G.~Jona-Lasinio, \PR{122} (1961) 345. A.~Casher,
\PLB{83} (1979) 395. T.~Banks and A.~Casher, \NPB{169} (1980) 103.

\end{list}

\end{document}